\documentclass[a4paper,manyauthors,nocleardouble,COMPASS]{cernphprep}

\usepackage{graphicx}
\usepackage{subfig}
\usepackage[percent]{overpic}
\usepackage[numbers, square, comma, sort&compress]{natbib}

\usepackage{ifpdf}

\usepackage[english]{babel}

\usepackage{graphicx}
\usepackage{rotating}
\usepackage{multirow}
\usepackage{amsmath}
\usepackage{array}
\usepackage{color}
\usepackage{ulem}
\usepackage{hyperref}

\input myunits.sty

\def\lapproxeq{\lower .7ex\hbox{$\;\stackrel{\textstyle<}{\sim}\;$}}
\def\gapproxeq{\lower .7ex\hbox{$\;\stackrel{\textstyle>}{\sim}\;$}}

\begin{document}
\selectlanguage{english}

\begin{titlepage}
\PHnumber{2016--xxx}
\PHdate{\today}

\vspace{1cm}
\title{Longitudinal double-spin asymmetry $A_1^{\rm p}$ \\and  spin-dependent structure             
function $g_1^{\rm p}$ of the proton\\                                          
         at {small} values of $x$ and $Q^2$}

\Collaboration{The COMPASS Collaboration}

\ShortAuthor{The COMPASS Collaboration}

\begin{abstract}
We present a precise
measurement of the proton longitudinal double-spin
asymmetry $A_1^{\rm p}$ and the proton spin-dependent structure
function $g_1^{\rm p}$ at photon virtualities $0.006\,(\GeV/c)^2<Q^2 < 1\,(\GeV/c)^2$ {in} the 
Bjorken $x$ range of $4 \times 10^{-5} < x < 4 \times 10^{-2}$.
The results are based on data collected by the
COMPASS Collaboration at CERN using muon beam
energies of $160\,\GeV$ and $200\,\GeV$. The
statistical precision is more than tenfold better than that of the previous
measurement in this region. In the whole range of $x$, the measured values of
$A_1^{\rm p}$ and $g_1^{\rm p}$ are found to be positive.
It is for the first time that spin effects are found at such low values of $x$.

\vspace*{60pt}
\end{abstract}

\vspace{80pt}
\Submitted{(to be submitted to Phys.\ Lett.\ B)}
\vfill

%
%
\section*{The COMPASS Collaboration}
\label{app:collab}
\renewcommand\labelenumi{\textsuperscript{\theenumi}~}
\renewcommand\theenumi{\arabic{enumi}}
\begin{flushleft}
M.~Aghasyan\Irefn{triest_i},
M.G.~Alexeev\Irefn{turin_u},
G.D.~Alexeev\Irefn{dubna}, 
A.~Amoroso\Irefnn{turin_u}{turin_i},
V.~Andrieux\Irefnn{illinois}{saclay},
N.V.~Anfimov\Irefn{dubna}, 
V.~Anosov\Irefn{dubna}, 
A.~Antoshkin\Irefn{dubna}, 
K.~Augsten\Irefnn{dubna}{praguectu}, 
W.~Augustyniak\Irefn{warsaw}
% ...
\end{flushleft}
%
%
\begin{Authlist}
\item \Idef{aveiro}{University of Aveiro, Dept.\ of Physics, 3810-193 Aveiro, Portugal}
\item \Idef{bochum}{Universit\"at Bochum, Institut f\"ur Experimentalphysik, 44780 Bochum, Germany\Arefs{l}\Aref{s}}
\item \Idef{bonniskp}{Universit\"at Bonn, Helmholtz-Institut f\"ur  Strahlen- und Kernphysik, 53115 Bonn, Germany\Arefs{l}}
\item \Idef{bonnpi}{Universit\"at Bonn, Physikalisches Institut, 53115 Bonn, Germany\Arefs{l}}
\item \Idef{brno}{Institute of Scientific Instruments, AS CR, 61264 Brno, Czech Republic\Arefs{m}}
\item \Idef{calcutta}{Matrivani Institute of Experimental Research \& Education, Calcutta-700 030, India\Arefs{n}}
\item \Idef{dubna}{Joint Institute for Nuclear Research, 141980 Dubna, Moscow region, Russia\Arefs{o}}
\item \Idef{freiburg}{Universit\"at Freiburg, Physikalisches Institut, 79104 Freiburg, Germany\Arefs{l}\Aref{s}}
\item \Idef{cern}{CERN, 1211 Geneva 23, Switzerland}
\item \Idef{liberec}{Technical University in Liberec, 46117 Liberec, Czech Republic\Arefs{m}}
\item \Idef{lisbon}{LIP, 1000-149 Lisbon, Portugal\Arefs{p}}
\item \Idef{mainz}{Universit\"at Mainz, Institut f\"ur Kernphysik, 55099 Mainz, Germany\Arefs{l}}
\item \Idef{miyazaki}{University of Miyazaki, Miyazaki 889-2192, Japan\Arefs{q}}
\item \Idef{moscowlpi}{Lebedev Physical Institute, 119991 Moscow, Russia}
\item \Idef{munichtu}{Technische Universit\"at M\"unchen, Physik Dept., 85748 Garching, Germany\Arefs{l}\Aref{r}}
\item \Idef{nagoya}{Nagoya University, 464 Nagoya, Japan\Arefs{q}}
\item \Idef{praguecu}{Charles University in Prague, Faculty of Mathematics and Physics, 18000 Prague, Czech Republic\Arefs{m}}
\item \Idef{praguectu}{Czech Technical University in Prague, 16636 Prague, Czech Republic\Arefs{m}}
\item \Idef{protvino}{State Scientific Center Institute for High Energy Physics of National Research Center `Kurchatov Institute', 142281 Protvino, Russia}
\item \Idef{saclay}{IRFU, CEA, Universit\'e Paris-Saclay, 91191 Gif-sur-Yvette, France\Arefs{s}}
\item \Idef{taipei}{Academia Sinica, Institute of Physics, Taipei 11529, Taiwan\Arefs{tw}}
\item \Idef{telaviv}{Tel Aviv University, School of Physics and Astronomy, 69978 Tel Aviv, Israel\Arefs{t}}
\item \Idef{triest_u}{University of Trieste, Dept.\ of Physics, 34127 Trieste, Italy}
\item \Idef{triest_i}{Trieste Section of INFN, 34127 Trieste, Italy}
\item \Idef{turin_u}{University of Turin, Dept.\ of Physics, 10125 Turin, Italy}
\item \Idef{turin_i}{Torino Section of INFN, 10125 Turin, Italy}
\item \Idef{tomsk}{Tomsk Polytechnic University,634050 Tomsk, Russia\Arefs{nauka}}
\item \Idef{illinois}{University of Illinois at Urbana-Champaign, Dept.\ of Physics, Urbana, IL 61801-3080, USA\Arefs{nsf}}
\item \Idef{warsaw}{National Centre for Nuclear Research, 00-681 Warsaw, Poland\Arefs{u}}
\item \Idef{warsawu}{University of Warsaw, Faculty of Physics, 02-093 Warsaw, Poland\Arefs{u}}
\item \Idef{warsawtu}{Warsaw University of Technology, Institute of Radioelectronics, 00-665 Warsaw, Poland\Arefs{u} }
\item \Idef{yamagata}{Yamagata University, Yamagata 992-8510, Japan\Arefs{q} }
\end{Authlist}
%
%
\renewcommand\theenumi{\alph{enumi}}
\begin{Authlist}
\item [{\makebox[2mm][l]{\textsuperscript{\#}}}] Corresponding authors
\item [{\makebox[2mm][l]{\textsuperscript{*}}}] Deceased
\item \Adef{a}{Also at Instituto Superior T\'ecnico, Universidade de Lisboa, Lisbon, Portugal}
\item \Adef{b}{Also at Dept.\ of Physics, Pusan National University, Busan 609-735, Republic of Korea and at Physics Dept., Brookhaven National Laboratory, Upton, NY 11973, USA}
\item \Adef{ictp}{Also at Abdus Salam ICTP, 34151 Trieste, Italy}
\item \Adef{r}{Supported by the DFG cluster of excellence `Origin and Structure of the Universe' (www.universe-cluster.de) (Germany)}
\item \Adef{p2i}{Supported by the Laboratoire d'excellence P2IO (France)}
\item \Adef{d}{Also at Chubu University, Kasugai, Aichi 487-8501, Japan\Arefs{q}}
\item \Adef{x}{Also at Dept.\ of Physics, National Central University, 300 Jhongda Road, Jhongli 32001, Taiwan}
\item \Adef{e}{Also at KEK, 1-1 Oho, Tsukuba, Ibaraki 305-0801, Japan}
\item \Adef{g}{Also at Moscow Institute of Physics and Technology, Moscow Region, 141700, Russia}
\item \Adef{h}{Present address: RWTH Aachen University, III.\ Physikalisches Institut, 52056 Aachen, Germany}
\item \Adef{yerevan}{Also at Yerevan Physics Institute, Alikhanian Br. Street, Yerevan, Armenia, 0036}
\item \Adef{y}{Also at Dept.\ of Physics, National Kaohsiung Normal University, Kaohsiung County 824, Taiwan}
\item \Adef{regensburg}{Also at Institut f\"ur Theoretische Physik, Universit\"at T\"ubingen, 72076 T\"ubingen, Germany}
\item \Adef{turin_p}{Also at University of Eastern Piedmont, 15100 Alessandria, Italy}
\item \Adef{i}{Present address: Uppsala University, Box 516, 75120 Uppsala, Sweden}
\item \Adef{c}{    Supported by the DFG Research Training Group Programmes 1102 and 2044 (Germany)} 
%
%
\item \Adef{l}{    Supported by BMBF - Bundesministerium f\"ur Bildung und Forschung (Germany)}
\item \Adef{s}{    Supported by FP7, HadronPhysics3, Grant 283286 (European Union)}
\item \Adef{m}{    Supported by MEYS, Grant LG13031 (Czech Republic)}
\item \Adef{n}{    Supported by SAIL (CSR) and B.Sen fund (India)}
\item \Adef{o}{    Supported by CERN-RFBR Grant 12-02-91500}
\item \Adef{p}{\raggedright 
                   Supported by FCT - Funda\c{c}\~{a}o para a Ci\^{e}ncia e Tecnologia, COMPETE and QREN, Grants CERN/FP 116376/2010, 123600/2011 
                   and CERN/FIS-NUC/0017/2015 (Portugal)}
\item \Adef{q}{    Supported by MEXT and JSPS, Grants 18002006, 20540299, 18540281 and 26247032, the Daiko and Yamada Foundations (Japan)}
\item \Adef{tw}{   Supported by the Ministry of Science and Technology (Taiwan)}
\item \Adef{t}{    Supported by the Israel Academy of Sciences and Humanities (Israel)}
\item \Adef{nauka}{Supported by the Russian Federation  program ``Nauka'' (Contract No. 0.1764.GZB.2017) (Russia)}
\item \Adef{nsf}{  Supported by the National Science Foundation, Grant no. PHY-1506416 (USA)}
\item \Adef{u}{    Supported by NCN, Grant 2015/18/M/ST2/00550 (Poland)}
\end{Authlist}

\end{titlepage}

\newpage
\setcounter{page}{1}

\setcounter{footnote}{0}
\section{Introduction}
\label{sec:intro}

The spin-dependent structure function of the proton, $g_1^{\rm p}$,
has been extensively studied in the last few decades. 
Precise measurements of $g_1^{\rm p}(x,Q^2)$
were realised in the deep inelastic regime of charged lepton nucleon scattering
at photon virtualities $Q^2 > 1\,(\GeV/c)^2$ \cite{aidala,compass_proton}
over a wide range of the Bjorken scaling variable $x$.
On the contrary, the behaviour of $g_1^{\rm p}$
at lower $Q^2$ is largely unknown. 
{For fixed-target experiments, the values of 
$Q^2\lapproxeq 1\,(\GeV/c)^2$ {imply small} values of $x$.
This low-$Q^2$ region is governed by `soft' processes and the 
transition to the region of higher $Q^2$ is still not understood.}

Quantum Chromodynamics (QCD) allows for a description of
`hard' interactions using a perturbative expansion
that is known to be applicable for $Q^2$ values
as low as about $1\,(\GeV/c)^2$. For lower values of $Q^2$, soft
interactions become relevant
and `non-perturbative' mechanisms dominate the {reaction} dynamics.
In order to provide a suitable description of the non-perturbative region
and also of the transition region between `soft' and `hard'
physics, it is tried in phenomenological calculations
to extrapolate ideas {based on the parton model} to the low-$Q^2$ region
and add mechanisms like (generalised) vector meson dominance, (G)VMD,
supplemented by the Regge model {(see Refs.~\cite{nonpert1,nonpert2,greco,chinese})}. 
New and precise data on
$g_1^{\rm p}(x,Q^2)$ in the low-$Q^2$ region are hence essential to
improve and validate such calculations.

Measurements at low $x$ and low $Q^2$ are scarce as they 
put very high demands on event triggering and reconstruction. 
In spin-dependent {lepto}production  they were performed
only by the Spin Muon Collaboration (SMC) 
using proton and deuteron targets \cite {smc_lowx}  
and by the COMPASS Collaboration using a deuteron target 
\cite{compass_g1d_lowx}. 
The latter, very precise results do not reveal any spin effects
in $g_1^{\rm d}$ {over} the whole measured interval of $x$. 
{In this Letter,}
we present new results obtained on the longitudinal double-spin 
asymmetry $A_1^{\rm p}$ and the spin-dependent structure function $g_1^{\rm p}$
for the proton, in the kinematic region $0.0062\,(\GeV/c)^2 < Q^2 < 1\,(\GeV/c)^2$ and
$4 \times 10^{-5} < x <  4 \times 10^{-2}$.  The data are analysed in four 2-dimensional grids of kinematic
variables, i.e. $(x,Q^2), (\nu, Q^2), (x, \nu)$ and $(Q^2,x)$, where $\nu$ denotes
the virtual-photon energy {in the target rest frame}. 
Note that the last grid differs from the first one in the number of bins 
chosen per variable.
The lower limit in $x$ coincides with that used in the
{COMPASS} low{-}$Q^2$ deuteron analysis \cite{compass_g1d_lowx}.
The low-$Q^2$ results presented in this Letter
complement our published proton  measurements covering {the high-$Q^2$ region}~\cite{compass_proton,g1p2010}.

{This Letter is organised as follows. We briefly describe the}
experimental set-up  
in Sec.~\ref{sec:experimen}, {the} event selection in Sec.~\ref{sec:data_sel} {and the} 
method of asymmetry calculation in Sec.~\ref{sec:asymm}. The results {are presented} in
Sec.~\ref{sec:results} and {the} summary is given in Sec.~\ref{sec:concl}.

\section{Experimental set-up}
\label{sec:experimen}
The measurements were performed using the COMPASS fixed-target set-up 
and positively charged muons provided by the M2 beam line of the CERN SPS.
{In 2007, the beam had} a momentum of
$160\,\,\GeV/c$ with ${5}\times 10^7$~$\mu^+$/s 
in $4.8\,\s$ long spills every {$16.8\,\s$}  and  
{in 2011 a momentum of} $200\,\,\GeV/c$ with  $10^7$~$\mu^+/\s$ 
in $10\,\s$ long spills every $40\,\s$.
The beam had a momentum spread of 5\%.
It was naturally polarised with a polarisation $P_{\rm b}$ of about {$-0.8$},
which is known with a precision of 5\%. Momentum and trajectory of
each incident muon were measured {before the target} 
by scintillator hodoscopes, scintillating fibre and silicon microstrip detectors.

A large solid-state target of ammonia (NH$_3$) inside a large-aperture 
superconducting solenoid provided longitudinally polarised protons. 
The proton polarisation was achieved by dynamic nuclear 
polarisation and reached a{n average} value of {$|P_{\rm t}|{\approx} 0.85$}. 
{The dilution factor $f$,  which accounts for the presence of unpolarisable material in the target,
is about 0.16 for ammonia.} 
The target material  was contained in three cylindrical cells of 4\,cm diameter
with  30\,cm, 60\,cm and 30\,cm length, which were 
separated by 5\,cm gaps and located along the beam one after the other. 
Neighbouring cells were polarised in opposite directions in 
order to use both target polarisations simultaneously during data taking. 
The polarisation directions were 
inverted on a regular basis by rotating the
direction of the target magnetic field, thus compensating for acceptance 
differences {between} different cells  
{and thereby minimising} possible systematic effects.
Once per year the direction of the polarisation with respect to the solenoid field 
was {reverted} by repolarisation in opposite direction 
keeping the solenoid field unchanged.
Ten NMR coils surrounding the target material allowed for a measurement
of $P_{\rm t}$ with a precision of 2\% in 2007 and 3.5\% in 2011. 

Momentum and angle of scattered muons and other produced particles {were} measured
in a two-stage open forward spectrometer with large angle and momentum 
acceptance using two dipole magnets with tracking detectors upstream and 
downstream of the magnets.
Scintillating fibre and micropattern gaseous detectors 
{were} employed in and 
close to the beam region, while multiwire proportional chambers, drift 
chambers and straw tube detectors cover{ed} the outer areas.

Scattered muons {were} identified by drift tube planes behind iron and concrete 
absorbers in both first and second stage of the spectrometer. Particle 
identification is not used in the current analysis.
Two different types of triggers {were} {employed}. 
``Inclusive'' triggers {were} based on 
coincidences of hodoscope signals produced by scattered muons.
``Semi-inclusive'' triggers require{d} an energy deposit in one of 
the calorimeters with an optional coincidence with an inclusive trigger.
The reader is referred to Ref.~\cite{nimpaper} 
for the detailed description of the muon beam,    
the three-cell polarised NH$_3$ target and the COMPASS spectrometer.

\section{Event selection}
\label{sec:data_sel}

Events {selected for the analysis} are required to have a  reconstructed 
incoming muon, a scattered muon and an 
interaction vertex. As scattering angles
{in the laboratory frame} are very small for 
low-$Q^2$ events, at least one additional track attached to the vertex
is required to improve the vertex resolution {in beam direction}.
For the 2007 data, incoming muon momenta  are required to range between $140\,\GeV/c$ and 
$160\,\GeV/c$, and for the 2011 data  between $185\,\GeV/c$ and $215\,\GeV/c$.
In order to equalise the beam flux through all target cells, the extrapolated
track of the incoming muon is required to pass through all target cells.
{Interactions originating from the} unpolarised material 
surrounding the target {are rejected} by {imposing}
appropriate {constraints} on the {position of the} interaction vertex.
The scattered muon is identified by requiring that it {has} passed more than 15
radiation length{s of material} and it has to point {back} to the hodoscope 
that triggered the event.
Kinematic constraints are applied {on} the photon virtuality{,} 
$Q^2 <$ 1 (GeV/$c$)$^2$, and {on} the
{Bjorken} scaling variable,
$x > 4\times 10^{-5}$, as {it was done} in the analysis of the COMPASS deuteron {data~}\cite{compass_g1d_lowx}. 
The latter {constraint} is used to {avoid 
the region where $x$ cannot be determined with sufficient accuracy}.
In addition, the fraction of the energy lost by the incoming muon {has to} fulfill the condition 
$0.1<y<0.9$, where the {lower limit} removes badly reconstructed events and the 
{upper limit} removes events with large radiative corrections as well as low-momentum 
muons resulting from pion decay{-in-flight}. These kinematic constraints 
lead to a minimum value of about $5\,\GeV/c^2$ 
{for $W$, where $W$ is the invariant mass of the 
$\gamma^* p$ system of virtual photon and proton.}

For a given primary interaction vertex with incident and scattered muon, we require 
at least one additional (hadron candidate) track that has to carry a fraction $z$
of the virtual photon energy with $0.1 < z < 1$ 
and a momentum $p < 140\,\GeV/c$ (2007) or $p < 180\,\GeV/c$ (2011).
{Here, t}he condition on $z$ rejects poorly reconstructed tracks {and} the condition
on $p$ removes beam halo muons.
This ``hadron method''~ \cite{smc} does not only
improve the resolution of the {primary} interaction vertex {but also} allows
{the reduction of} radiative background.

\begin{figure}[htb]
\begin{center}
\includegraphics[width=\textwidth,clip]{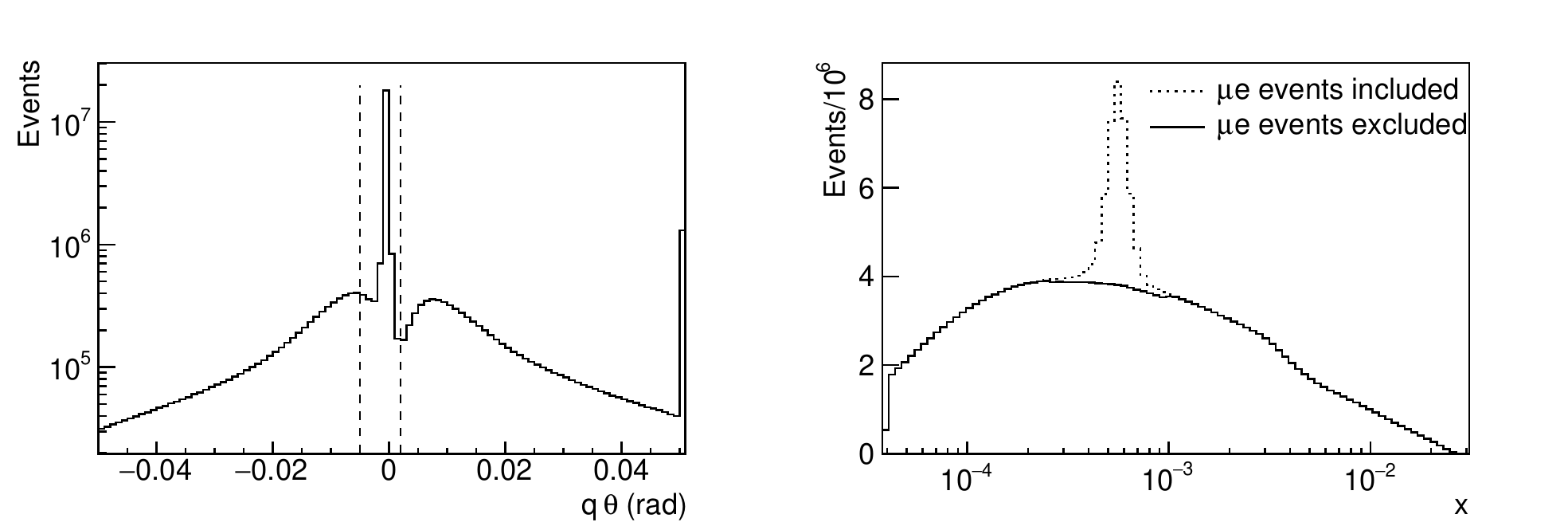}
\end{center}
\caption{\small 
{200~GeV data.} Left: Distribution of the variable $q\theta$ 
for events with one (positively or
negatively charged) additional track outgoing from the primary interaction vertex.
Events between vertical lines are removed from further analysis.
Note the logarithmic scale on the vertical axis.
Right: $x$ distribution of accepted events without and with 
$\mu e$  event rejection. 
}
\label{fig:mue}
\end{figure}

At the very low values of $x$ studied in this analysis, {there exists a 
contamination} by events {that originate from} elastic scattering of muons off 
atomic electrons {of} the target material. These events show up in the $x$ 
distribution as a prominent peak around the value 
$x_{\mu e}=m_\textrm{electron}/M=5.45\times 10^{-4}$, where $M$ 
is the proton mass. {We remove this contamination by imposing a constraint} 
on the product $q\theta$, where 
{$q=+1 \,(-1)$ is used if a particle of positive (negative) charge is}
associated to the track and {$\theta$ is} 
the angle between the hadron candidate track and the virtual-photon 
direction. In the range 
$-3.6<\log_{10}(x)<-3.0$, events {with one hadron candidate are} rejected if  
$-0.005\,\rad <q\theta < 0.002\,\rad$ {and} events with 
{two} hadron candidates if $-0.001\,\rad <q\theta < 0\,\rad$. 
{For the former case and either hadron charge, t}he distribution of the 
product $q\theta$ is presented in Fig.~\ref{fig:mue}~(left). {I}n Fig.~\ref{fig:mue}~(right),
the $x$ distributions of accepted events are shown {without and with the constraint} on  $q\theta$.

After having applied all selection criteria, 447 million events taken with a beam 
energy of $160\,\GeV$  and 229 million taken with $200\,\GeV$ remain for analysis.

\section{Asymmetry extraction}
\label{sec:asymm}

\begin{figure}[htbp]
\includegraphics[width=0.5\textwidth,clip]{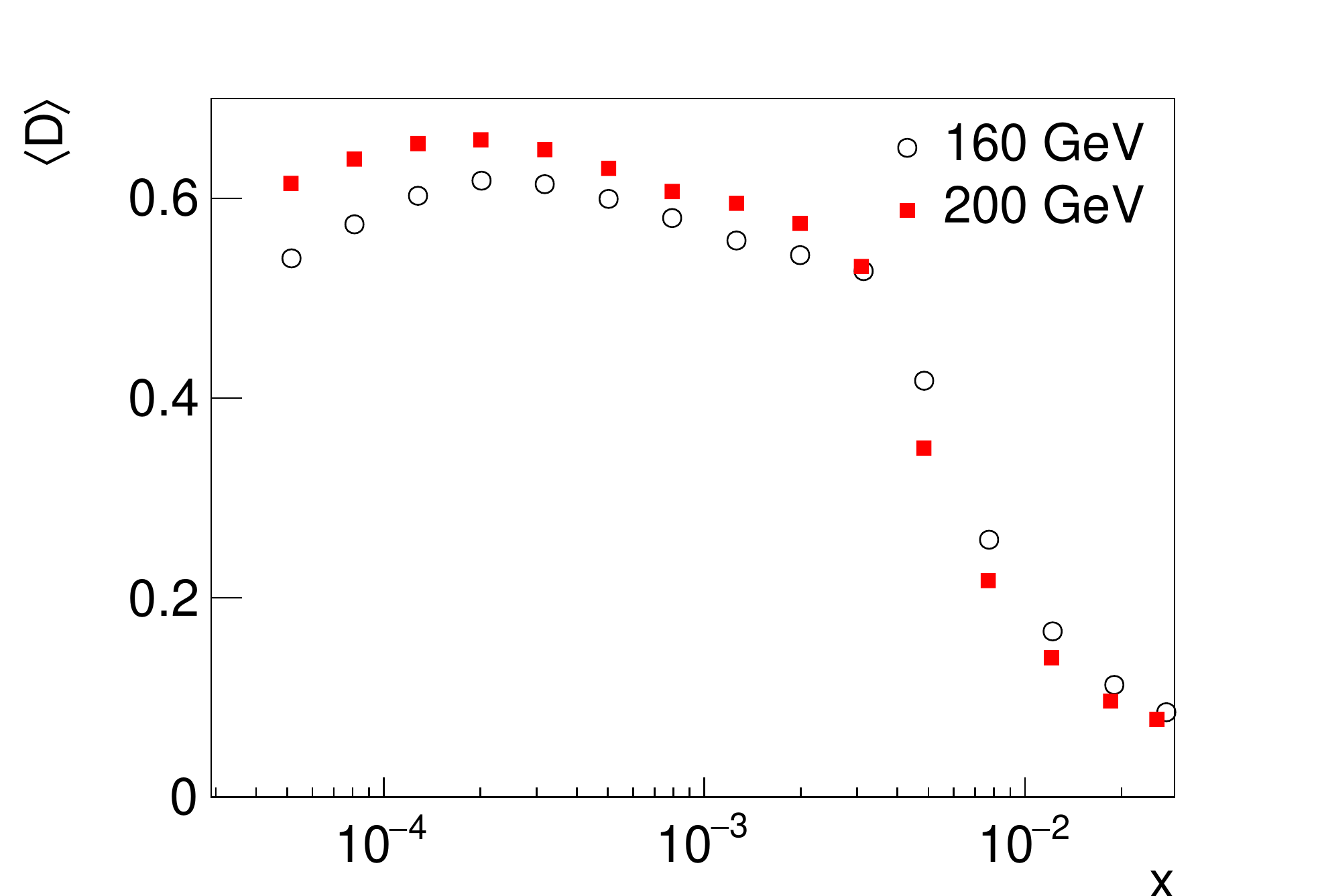}
\includegraphics[width=0.5\textwidth,clip]{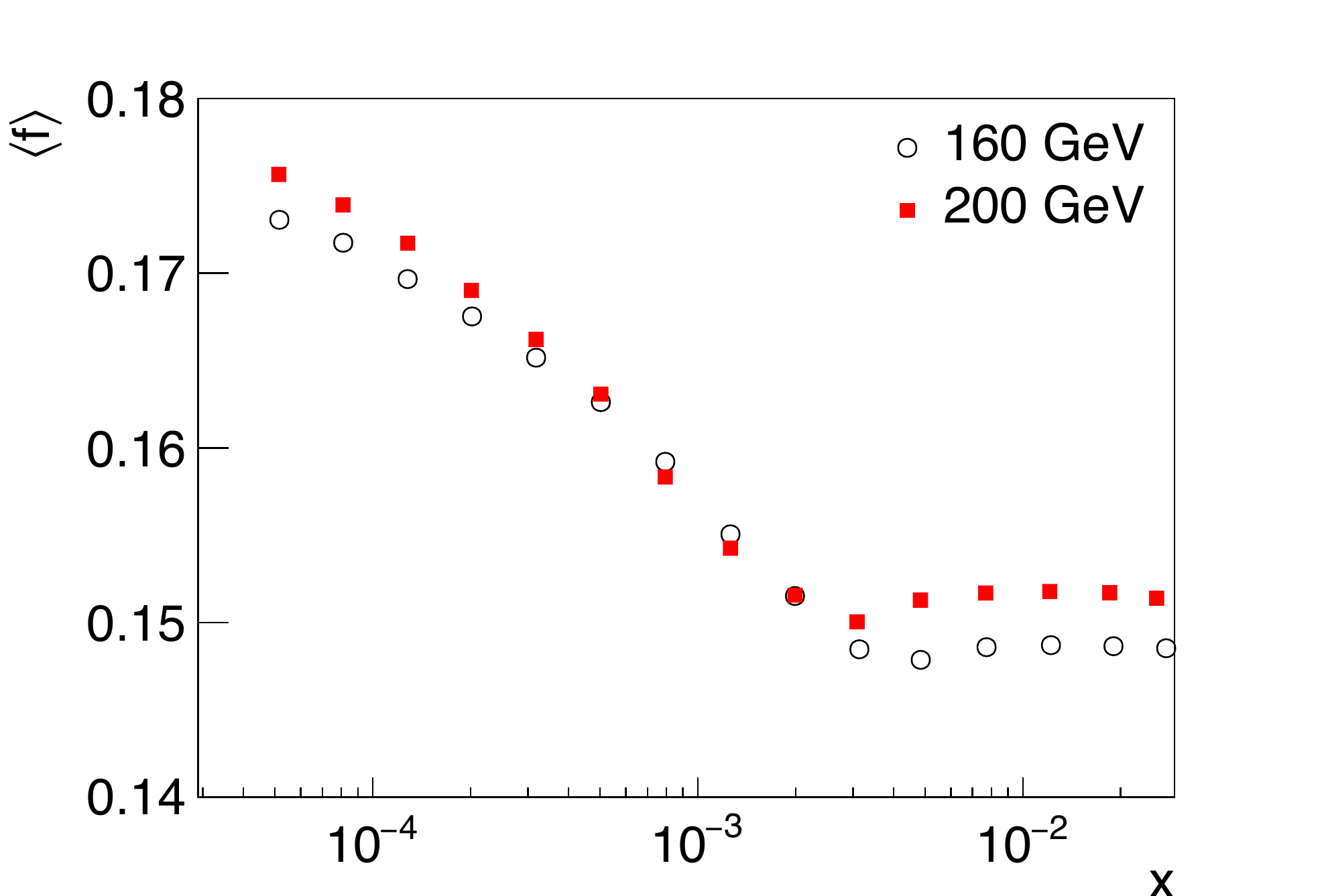}

\caption{\small Mean depolarisation factor (left) and  mean 
dilution factor (right)
as a function of $x$. 
}
    \label{fig:depol2_R2}
\end{figure}

The longitudinal double-spin lepton-proton cross-section asymmetry is given by 
\begin{equation}
A_\text{LL}^{\rm p}=
\frac{\sigma^{\stackrel{\rightarrow}{\Leftarrow}}-\sigma^{\stackrel{\rightarrow}{\Rightarrow}}}
{\sigma^{\stackrel{\rightarrow}{\Leftarrow}}+\sigma^{\stackrel{\rightarrow}{\Rightarrow}}}
=D(A_1^{\rm p}+\eta A_2^{\rm p}),
\label{all}
\end{equation}
where the arrows refer to the longitudinal spin orientations of incoming muon ($\rightarrow$)
and target proton ($\Rightarrow$). 
It can be decomposed into a longitudinal photon-nucleon asymmetry $A_1^{\rm p}$ and a transverse photon-nucleon asymmetry $A_2^{\rm p}$, where the longitudinal asymmetry is defined in terms of 
the $\gamma^*$p cross sections as
\begin{equation}
A_1^{\rm p}=\frac{\sigma_{1/2}-\sigma_{3/2}}{\sigma_{1/2}+\sigma_{3/2}}.
\end{equation}
Here, the subscript refers to the total angular momentum 
of the $\gamma^*$p system.
The factor $D$ in Eq.~(\ref{all}) is the so-called depolarisation factor and $\eta$ is a 
kinematic factor. 
Full expressions for $D$ and $\eta$ 
are given in Ref. \cite{compass_g1d_lowx}; the behaviour of $D$ in the
kinematic region of this analysis is shown in 
Fig.~\ref{fig:depol2_R2}~(left). 
In the COMPASS kinematic range, the {factor} 
$\eta$ is negligible, hence  the term containing the transverse asymmetry  $A_2^{\rm p}$ is 
{of negligible size} and its possible contribution is included in the systematic uncertainty 
of $A_1^{\rm p}$. 

The number of events originating from a given target cell with a given direction of the
target polarisation can be expressed as
\begin{equation}
N_i = a_i \phi_i n_i \bar{\sigma}(1+P_{\rm b}P_{\rm t} f D A_1^{\rm p}),~~~i=o1,c1,o2,c2.  
\label{4}
\end{equation}
Here, $a_i$ is the acceptance, $\phi_i$ the beam flux, $n_i$ 
the number of target nuclei,  $\bar\sigma$ the spin-independent 
cross section and $f$ the dilution factor.
{The four equations of Eq.~(\ref{4}) denoted by {the subscript} 
$i$ are giving the numbers of events that originate from either the combined
outer cells ($o$) or the central cell ($c$), each for the two directions of the
solenoid field (1 or 2).}
They are combined into a second-order equation in 
$A_1^{\rm p}$ for the ratio $(N_{o1}N_{c2})/(N_{o2}N_{c1})$, where the product 
$a_i \phi_i n_i \bar{\sigma}$ cancels provided that the 
ratio of acceptances of the central cell $c$ and the outer cells $o$ is the same 
before and after field reversal.
In order to minimise the statistical uncertainty  of the asymmetry,
the events are weighted by a factor $w=P_{\rm b}fD$. The target polarisation
is not included in the weight because it is
not constant over time and would hence generate false asymmetries.

The dilution factor $f$ includes a correction
factor $\rho = \sigma^{\rm p}_{1\gamma}/\sigma^{\rm p}_{\rm tot}$ \cite{terad} that accounts
for radiative events originating from  unpolarised target protons.
This effective dilution factor depends only weakly on the incident energy, 
decreases with $x$ and reaches a value 
of about 0.17 at $x\sim 10^{-4}$ and about 0.15 at $x\sim 10^{-2}$.
Its relative uncertainty amounts to 5\%.
The $x$ dependence of the average dilution factor is shown in 
Fig.~\ref{fig:depol2_R2} (right).
The beam polarisation is a function of the beam momentum
and is taken from a parametrisation based on a Monte Carlo
simulation of the beam line, which was validated by SMC \cite{beampol}.

The final value of $A_1^{\rm p}$ is obtained as weighted average
of the values calculated for the two target-spin orientations.
{I}t is corrected for spin-dependent radiative 
effects~\cite{polrad} and for the polarisation of the $^{14}$N nuclei
present in the target. It was verified that the use of semi-inclusive 
triggers and the requirement of a reconstructed hadron 
do not bias the determination of $A_1^{\rm p}$~{\cite{smc,malte_phd}}. 
More details on the analysis can be found in Ref.~\cite{asn_phd}.

The {additive and multiplicative systematic uncertainties of $A_1^{\rm p}$
are} 
shown in Table~\ref{tab:sys_error}.
The largest multiplicative contribution
originates from the depolarisation factor $D$ through its dependence
on the poorly known function $R=\sigma_{\rm L}/\sigma_{\rm T}$, which is the 
ratio of the absorption cross sections 
of longitudinally and transversely  polarised virtual photons.
The parameterisation of the function $R$ described in detail in
Ref. \cite{compass_g1d_lowx} takes into account all existing
measurements together with an extension to very low values of $Q^2$. 
As systematic uncertainty of $R$ a constant value of 0.2 
is taken for $Q^2<0.2\,(\GeV/c)^2$. 
The largest additive contribution
originates from possible false asymmetries, which are estimated from time-dependent
instabilities in the spectrometer as described in Ref.~\cite{asn_phd}. In certain 
kinematic regions, it can be larger than the statistical uncertainty.
\begin{table}[h!]
\caption{\small Systematic uncertainties of $A_1^{\rm p}$ and $g_1^{\rm p}$.}
\begin{center}
\renewcommand{\arraystretch}{1.2}

\begin{tabular}{|l|l|c|c|c|}
\hline
\hline
& & & $\Delta A_1^{\rm p}$&$\Delta g_1^{\rm p}$\\
\hline
               & Beam polarisation     & $\Delta P_{\rm b}/P_{\rm b}$ & 5\% & 5\%\\
\cline{2-5}
Multiplicative & Target polarisation   & $\Delta P_{\rm t}/P_{\rm t}$ & 2\% (2007)& 2\% (2007)\\
&&&  3.5\% (2011) & 3.5\% (2011)\\
\cline{2-5}
contribution      & Depolarisation factor & $\Delta D/D$ & 4 -- 32 \% & --\\
\cline{2-5}
& Dilution factor & $\Delta f/f$    & 5\% & 5\%\\
\cline{2-5}
& $D(1+R)$ & $\Delta(D(1+R))/D(1+R)$    & --& 0.02\% -- 6\%\\
\cline{2-5}
& $F_2$ & $\Delta F_2/F_2$    & --& 7\% -- 31\%\\
\hline
\hline
Additive       & Transverse asymmetry  & ${(\eta/\rho) \Delta A_2}$ & 
$<0.03\,\Delta A_1^{\rm stat}$ &$<0.03\,\Delta g_1^{\rm stat}$\\
\cline{2-5}
contribution      & Radiative corrections& $\Delta A_1^{RC}$ &
$< 0.02 \,\Delta A_1^{\rm stat}$&$< 0.02 \,\Delta g_1^{\rm stat}$\\
\cline{2-5}
 & False asymmetries   & $A_{\rm false}$ & $<1.3\,\Delta A_1^{\rm stat}$ &
$<1.3\,\Delta g_1^{\rm stat}$\\
\hline
\hline
\end{tabular}
\end{center}
\label{tab:sys_error}
\end{table}

The spin-dependent structure function of the proton, $g_1^{\rm p}$, is
determined from the virtual-photon asymmetry $A_1^{\rm p}$ neglecting 
$A_2^{\rm p}$:
\begin{equation}
g_1^{\rm p}=\frac{F_2^{\rm p}}{2 x (1+R)}A_1^{\rm p}.
\label{5}
\end{equation}
Here, $F_2^{\rm p}$ is the spin-independent structure function of the proton. 
For $F_2^{\rm p}$ we used 
{the SMC parameterisation~\cite{smc} within its validity limits, i.e.   
$x > 0.0009$ and $Q^2 > 0.2\,(\GeV/c)^2$. 
O}utside these limits, the values were calculated
using the phenomenological model of Refs.~\cite{KB1989,BK1992}, 
which is based on the GVMD concept. 
Equation~(\ref{5}) can be written as 
\begin{equation}
g_1^{\rm p}=\frac{F_2^{\rm p}}{2x\, D(1+R)}   A_{\rm LL}^{\rm p},
\label{6}
\end{equation}
so that the systematic uncertainty of $g_1^{\rm p}$ can be obtained from the 
following three components: i) the systematic
uncertainty of $A_{\rm LL}^{\rm p}\equiv A_1^{\rm p}/D$, ii) the systematic uncertainty of $F_2^{\rm p}$, 
and iii) the systematic uncertainty of the product $D(1+R)$. The systematic uncertainties of $A_{\rm LL}$
and $R$
were already discussed above. The systematic uncertainty of $F_2^{\rm p}$ is estimated 
from the difference between the SMC parameterisation 
and the models {of Refs.}~\cite{BGK2002,KB1989,ALLM97}.
It is taken as half of the maximum
of the absolute differences between the used parameterisation or  model and 
the remaining  models. 
For $Q^2>0.2\,(\GeV/c)^2$, this is always the absolute
value of the difference between the SMC parameterization and the model of 
Refs.~\cite{BGK2002,KB1989}.
When calculating $g_1^p$ using Eq.~(\ref{6}) instead of Eq.~(\ref{5}), we benefit from  the 
fact that $D$ and $R$
are correlated (see also Ref. \cite{compass_g1d_lowx}), which results in a
reduced systematic uncertainty compared to the one of $A_1^{\rm p}$.

\section{Results}
\label{sec:results}

We present here the results for the spin asymmetry $A_1^{\rm p}$ and the spin structure
function $g_1^{\rm p}$ measured in the kinematic range $Q^2 < 1~(\GeV/c)^2$ and
$4 \times 10^{-5}< x <ß 4 \times 10^{-2}$ using the two beam energies 
$160\,\GeV$ and $200\,\GeV$. 
{For each beam energy, the data are analysed in four 
two-dimensional grids: $(x,Q^2)$, $(\nu,Q^2)$, $(x,\nu)$
and $(Q^2,x)$, where the latter has a smaller number of $x$ bins.}

\begin{figure}[htbp]
\begin{center}
\includegraphics[width=0.49\textwidth,clip]{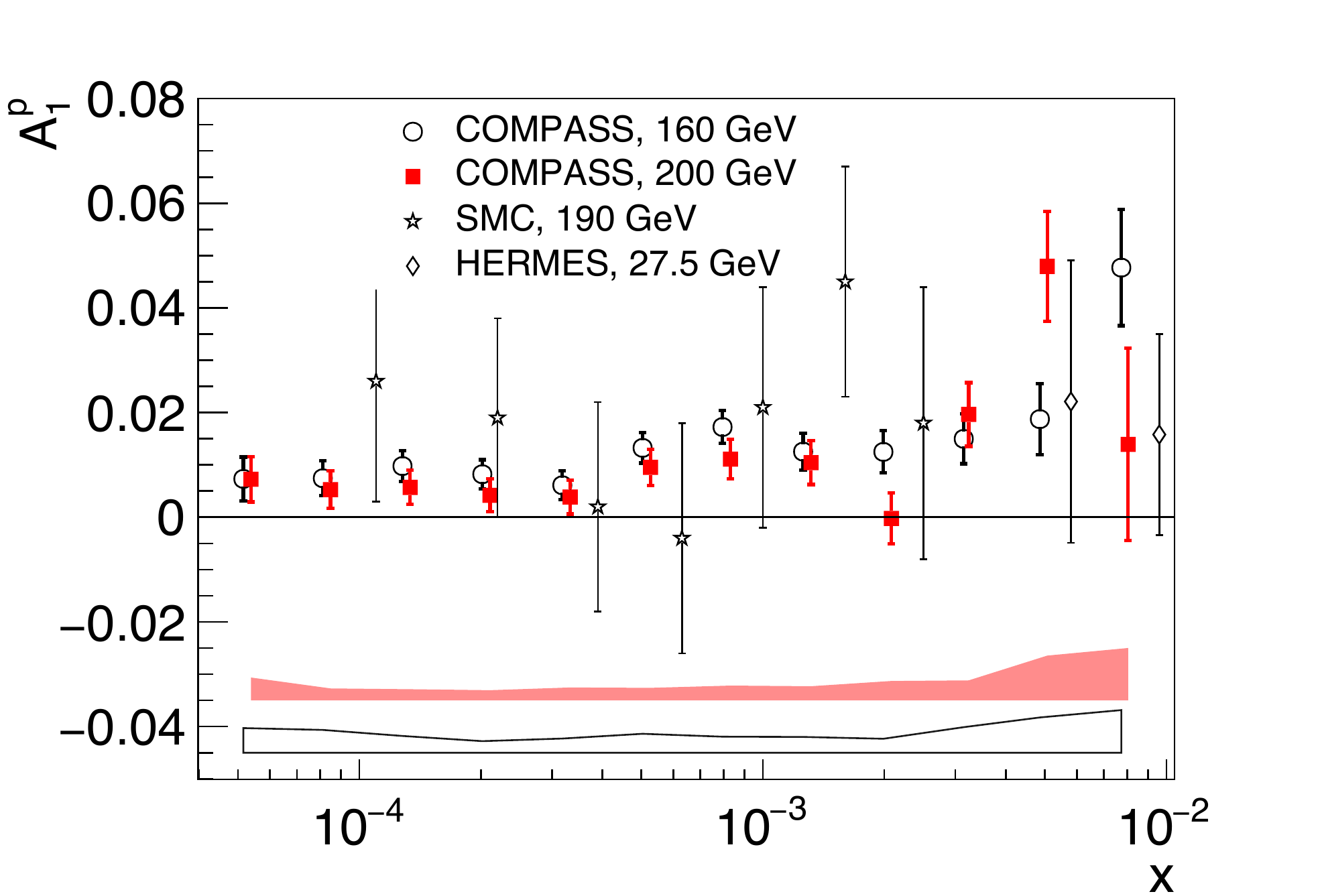}
\includegraphics[width=0.49\textwidth,clip]{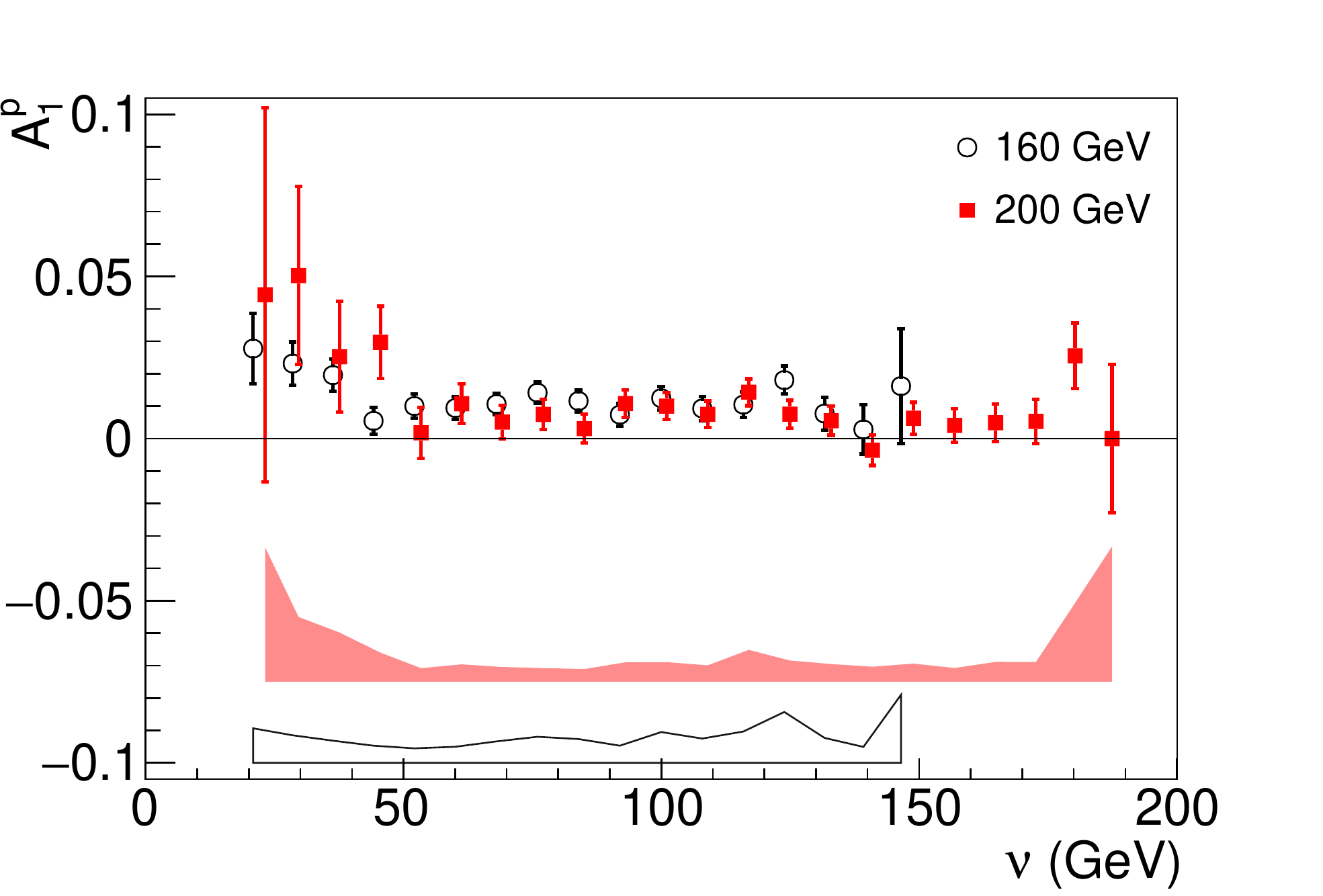}
\end{center}
\caption{\small The asymmetry $A_1^{\rm p}$ as a function 
of $x$ at the measured $Q^2$ values for $x<0.01$ (left) and as a function
of $\nu$ (right). 
Error bars represent statistical and bands systematic uncertainties.
On the left, results from other experiments \cite{smc_lowx,smc,hermes_new} 
are also shown.
}
\label{fig:a1_compar}
\end{figure}

\begin{figure}[htpb]
\begin{center}
\includegraphics[width=.9\textwidth,clip]{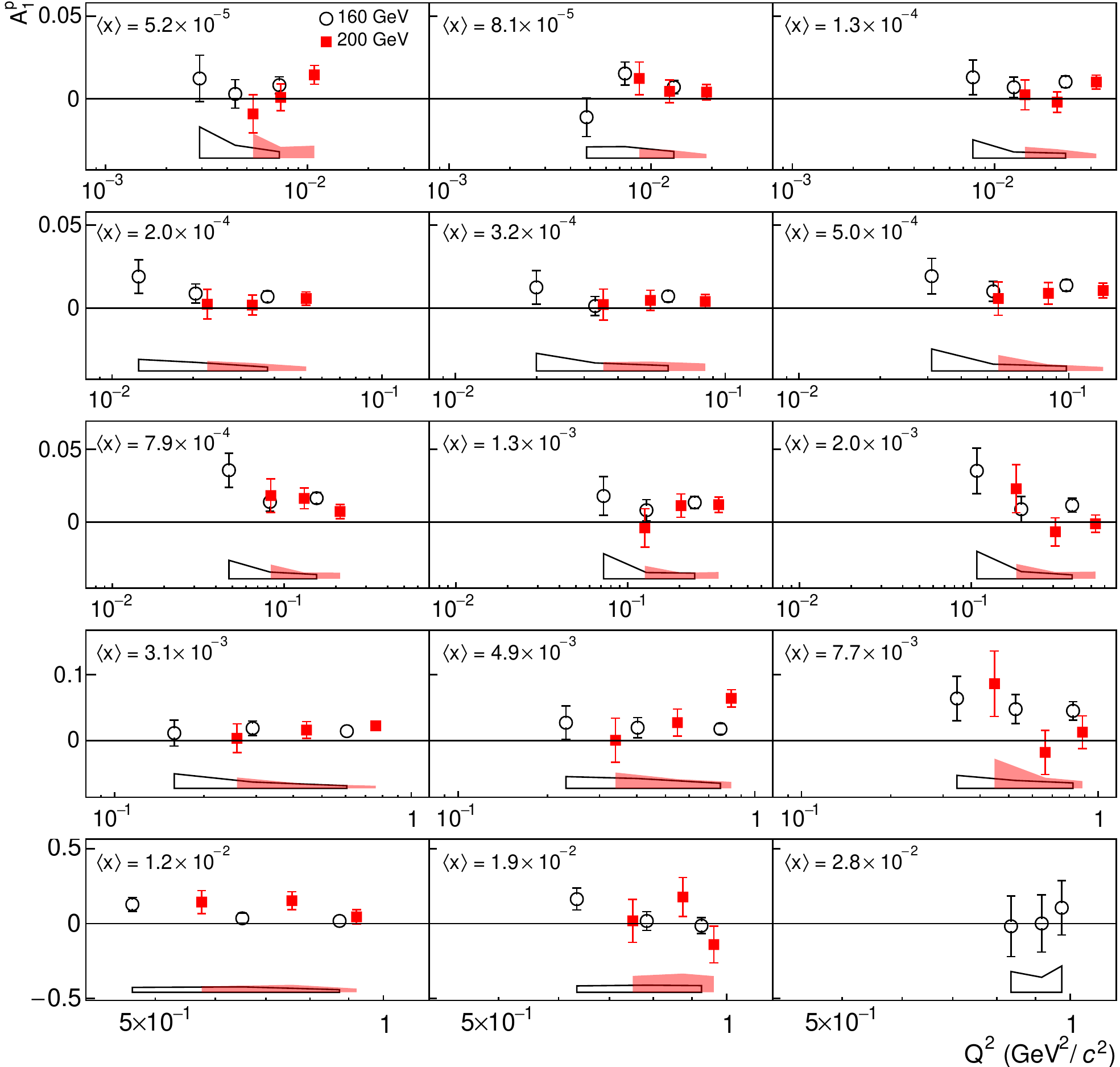}
\end{center}
\caption{\small The asymmetry $A_1^{\rm p}$ as a function of $Q^2$ in 
15 bins of $x$ for the two beam energies.
The bands indicate the size of the systematic uncertainties.
}
\label{fig:a1}
\end{figure}

The $x$ dependence of $A_1^{\rm p}$ at the measured values of $Q^2$ is shown in
Fig.\,\ref{fig:a1_compar}~(left) for the two beam energies. 
A positive asymmetry {is observed, which} slightly {rises} with
$x$. It amounts to about 0.01 at $x<10^{-3}$, 
{indicating} for the first time the existence of spin effects at such small 
values of $x$. Note that the COMPASS 
results for the deuteron~\cite{compass_g1d_lowx} show an asymmetry $A_1^{\rm d}$ 
compatible with zero. 
In Fig.\,\ref{fig:a1_compar}~(left),
also the results for $A_1^{\rm p}$ from  SMC \cite{smc_lowx,smc}
and HERMES \cite{hermes_new} are shown. 
Within the large statistical uncertainties,
their results are consistent with our present results, but also with  zero.
Compared to the results from SMC, {which is} the only other 
experiment {that covers} the low-$x$ region, 
we improve the statistics by a factor of about 150.
In Fig.\,\ref{fig:a1_compar}~(right), the $\nu$-dependence of
$A_1^{\rm p}$ is shown. A rather flat distribution is
measured, apart from a slight enhancement for $\nu < 50\,\GeV$ that 
corresponds to higher values of $Q^2$.
In Fig.~\ref{fig:a1},
the results for $A_1^{\rm p}$ are shown versus $Q^2$ for the 15 bins in $x$. 
The results obtained at 160~GeV and 
200~GeV are consistent in the overlapping $Q^2$ region. {From the
figure, no conclusion on a possible $Q^2$ dependence can be drawn.} 

\begin{figure}[htpb]
\begin{center}
\includegraphics[width=0.5\textwidth,clip]{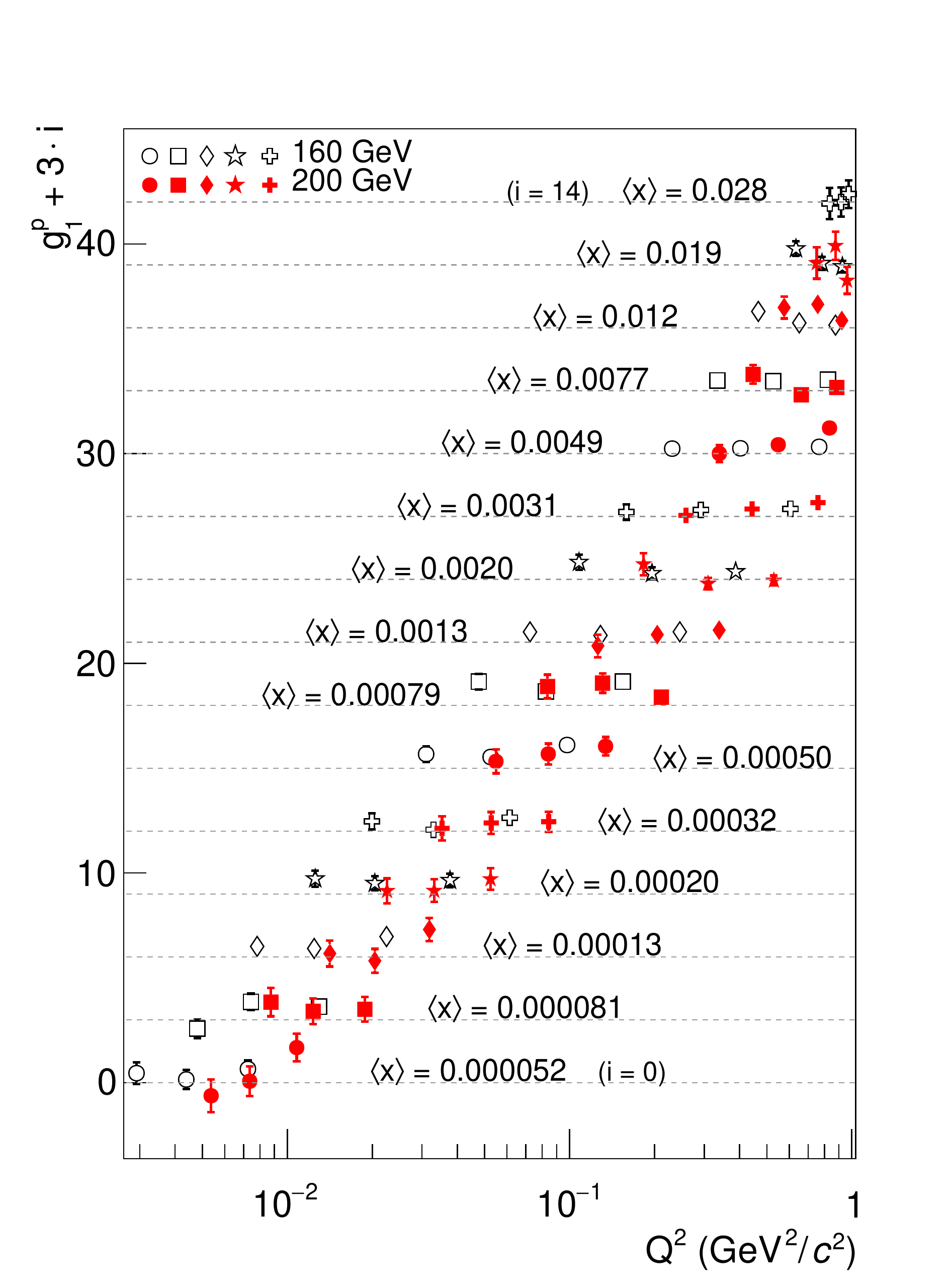}
\end{center}
\caption{\small The spin-dependent structure function  $g_1^{\rm p}$ as a function 
of $Q^2$ in 15 bins of $x$, shifted vertically for clarity. Closed (open) symbols correspond to 
160 GeV (200 GeV) data with {error bars showing} statistical uncertainties.
}
\label{fig:g1}
\end{figure}

\begin{figure}[htpb]
\begin{center}
\includegraphics[width=0.49\textwidth,clip]{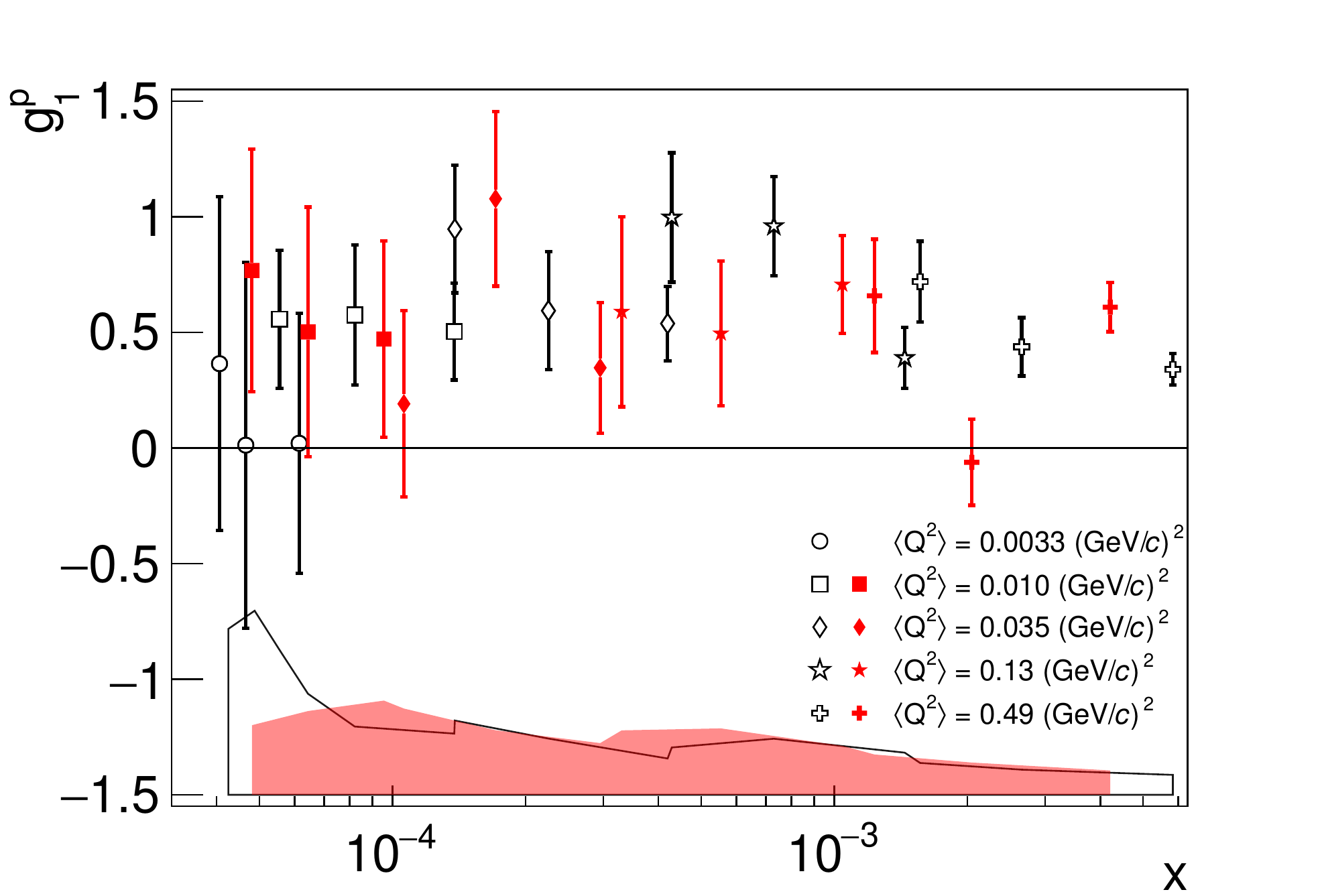}
\end{center}
\caption{\small The spin-dependent structure function  $g_1^{\rm p}$ as a function 
of $x$  in 5 bins of $Q^2$ . Closed (open) symbols correspond to 
160 GeV (200 GeV) data with 
error bars showing statistical uncertainties. Bands indicate the 
size of the systematic uncertainties.
The data points of the first bin in $Q^2$ are slightly shifted to the left 
for better visibility.
}
\label{fig:g1nostaggering}
\end{figure}

For the two beam energies, {our} results {on} $g_1^{\rm p}$ 
are shown versus $Q^2$ for the same 15 bins in $x$ (Fig.~\ref{fig:g1}) and
versus $x$ in 5 different bins in $Q^2$ (Fig.~\ref{fig:g1nostaggering}).
Down to the smallest value of $x$, i.e. $4 \times 10^{-5}$, $g_1^{\rm p}$ is 
positive within experimental uncertainties and does not show any 
trend to become negative or to grow with decreasing values of $x$.

All numerical values are available on HepData~\cite{hepdata}. 
The numerical values for $A_1^{\rm p}$ and $g_1^{\rm p}$
versus $x$, averaged over $Q^2$, are given together with their statistical 
and systematic uncertainites in Table~\ref{tab:a1x_g1a} of the appendix  
for the two energies 
separately. The data for the two energies were combined and
false asymmetries reevaluated for the merged data. The values for the combined
results are given in Table~\ref{tab:a1x_g1_combined}.

\begin{figure}[htpb]
\begin{center}
\includegraphics[width=0.49\textwidth,clip]{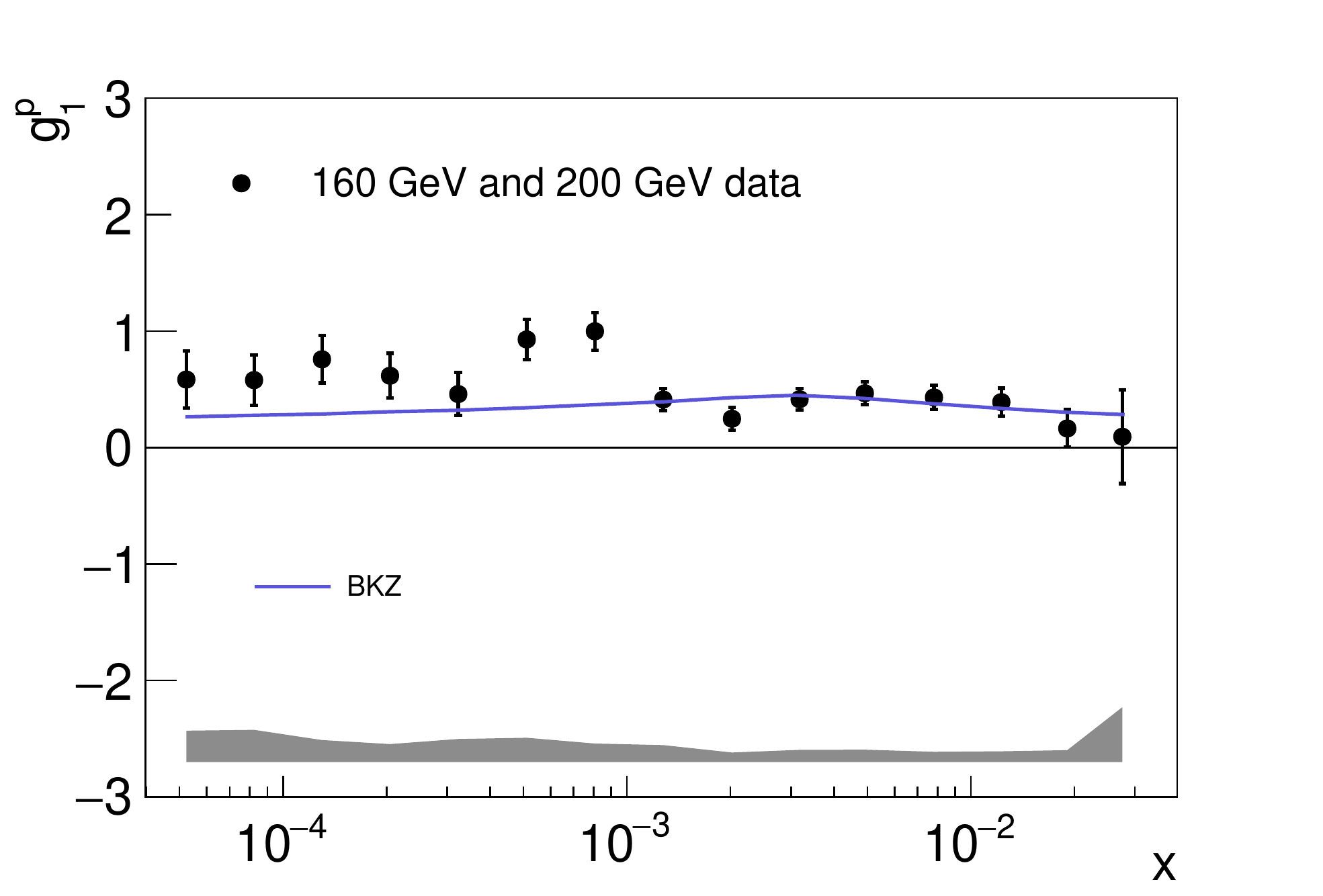}
\includegraphics[width=0.49\textwidth,clip]{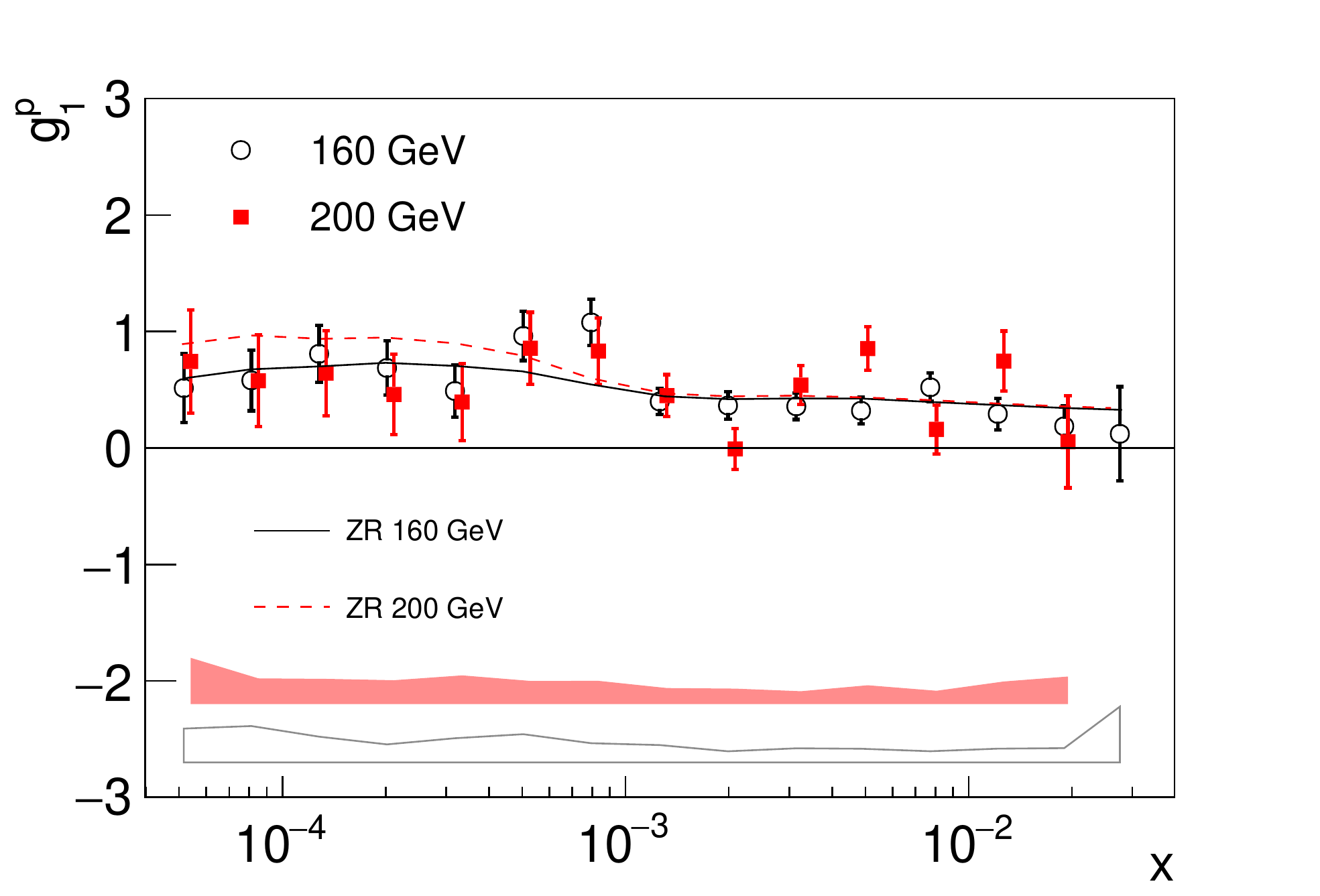}
\end{center}
\caption{\small 
Left: $x$ dependence of combined $g_1^{\rm p}$ data.
 The curve shows results of the $g_1^{\rm p}$ calculations of 
BKZ.~\cite{nonpert2}, where {for the parameterisation} of the  
perturbative part of $g_1^{\rm p}$
the DSSV~\cite{dssv} parton distributions at NLO accuracy. Right:
Comparison of the $x$ dependence of $g_1^{\rm p}$ at 160 GeV (open symbols) 
and 
200 GeV (closed symbols) with the results of the calculations of 
ZR~\cite{chinese} at 160 and 200 GeV incident energy (solid and dotted lines).
Error bars represent statistical and bands
systematic uncertainties.}
\label{fig:g1combined}
\end{figure}

In Fig.~\ref{fig:g1combined}, the present results on $g_1^{\rm p}$ are compared 
with the predictions of
the  phenomenological models of Refs.~\cite{nonpert2,chinese}.  
The first model (BKZ) is based on GVMD ideas supplemented by the 
Regge formalism. 
The contribution of heavy 
vector mesons to $g_1^{\rm p}$ was treated as an extrapolation of the 
QCD improved parton model to arbitrarily low values of $Q^2$.
The magnitude of the light vector meson contribution was 
fixed in the photoproduction limit by relating the first moment of $g_1^{\rm p}$ 
to the static properties of the proton via the Drell-Hearn-Gerasimov 
sum rule~\cite{dhg}, using the measurements in the region 
of baryonic resonances~\cite{baryon}. 
Both perturbative and non-perturbative contributions to $g_1^{\rm p}$ 
are found to be present and substantial at all values of $Q^2$.
Reasonable agreement is observed between the BKZ model
and our measurements in all four two-dimensional 
grids of kinematic variables. 
{Figure~\ref{fig:g1combined} (left) shows a comparison of the $x$~dependence of the BKZ model prediction with
the results for $g_1^{\rm p}$ obtained combining the 160~GeV and 200~GeV results.}
In the model of Ref.~\cite{chinese} (ZR), the nonperturbative part of
$g_1$ is also parameterised using the vector meson dominance mechanism 
together with Regge predictions (albeit done differently than in 
Ref.~\cite{nonpert2}),
while in the perturbative part QCD evolution {is employed} together with
parton recombination corrections. The $g_1^{\rm p}$ calculations
of Ref.~\cite{chinese} are presented in Fig.~\ref{fig:g1combined} (right), where
the broad bump at lowest values of $x$ is almost entirely 
due to the VMD contribution.

\section{Summary}
\label{sec:concl}

New results {are presented} on the longitudinal double-spin asymmetry $A_1^{\rm p}$ and 
the spin-dependent structure function $g_1^{\rm p}$ {of the proton}. 
In the kinematic domain {of the measurement,
$0.006~(\GeV/c)^2 < Q^2 < 1~(\GeV/c)^2$ and $4 \times 10^{-5} < x < 4 \times 10^{-2}$,
these results} improve the statistical
precision by a factor of more than 10 compared to existing measurements.

The values of $A_1^{\rm p}(x)$ and $g_1^{\rm p}(x)$ are found to be positive 
over the whole measured range of $x$, with a value of about 0.01 for the spin asymmetry for $x<10^{-3}$. 
While the earlier results obtained using a deuteron target were found to be
consistent with zero, the present measurement shows for the first time
non-zero spin effects at such small values of $x$. 
The data are compared to two phenomenological models of $g_1^{\rm p}$ valid in the region of low $x$ and 
low $Q^2$~\cite{nonpert2,chinese}, which are based on vector meson dominance 
and partonic ideas suitably 
extrapolated to low values of $Q^2$. These models describe the general trend in 
the data and 
indicate the existence of substantial perturbative and non-perturbative contributions to 
$g_1^{\rm p}$ in the whole $Q^2$ range of the data.

\section*{Acknowledgments}
We are grateful to J.~Ruan and W.~Zhu for discussions and supplying us with
the $g_1^{\rm p}$ values and to R.~Sassot and W.~Vogelsang of DSSV for 
supplying as with the
code to calculate their parton distributions and for their values of $g_1^{\rm p}$.
We gratefully acknowledge the support of the CERN management and staff
and the skill and effort of the technicians of our collaborating
institutes. This work was made possible 
by the financial support of our funding agencies.


\appendix
\setlength{\textheight}{53\baselineskip}
\section{Appendix}

The results for $A_1^{\rm p}$ and $g_1^{\rm p}$ for 160~GeV and 200~GeV are given
in Table~\ref{tab:a1x_g1a} and the combined results in Table~\ref{tab:a1x_g1_combined}.

\begin{table}[h!]
\begin{center}
{\footnotesize
\caption{\small
Values of $A_1^{\rm p}$ and $g_1^{\rm p}$ with their statistical and systematic uncertainties
  as a function of $x$ and the average values of $x$, $Q^2$ and $y$, 
for the 160~GeV and 200~GeV data.
  The maximum value of $Q^2$ used in the analysis is $1\,(\GeV/c)^2$. 
Bins in $x$ are of equal width in log$_{10}x$.}
\label{tab:a1x_g1a}
\begin{tabular}{|r@{--}l|c|c|c|c|c|}
\hline \hline
\multicolumn {2}{|c|}{}&&&&& \\
\multicolumn {2}{|c|}{$x$ range} & $\langle x \rangle$ & $\langle Q^2 \rangle $  & $\langle y\rangle $  & \multicolumn{1}{c|}{$A_1^{\rm p}$} & \multicolumn{1}{c|}{$g_1^{\rm p}$} \\
\multicolumn {2}{|c|}{}   &  &   [(GeV$/c$)$^2$]&  &  &  \\
\hline \hline
\multicolumn{7}{|c|}{160~GeV data}\\
\hline
  0.00004  &  0.000063 &$0.000052$&$0.0062$&$0.40$&$ 0.0073\pm  0.0042\pm 0.0047$&$ 0.51\pm 0.29\pm 0.29$\\  \hline
  0.000063 &  0.0001   &$0.000081$&$ 0.011$&$0.45$&$ 0.0074\pm  0.0034\pm 0.0044$&$ 0.58\pm 0.26\pm 0.31$\\  \hline
  0.0001   &  0.00016  &$ 0.00013$&$ 0.019$&$0.49$&$ 0.0098\pm  0.0029\pm 0.0032$&$ 0.81\pm 0.24\pm 0.22$\\  \hline
  0.00016  &  0.00025  &$ 0.00020$&$ 0.032$&$0.53$&$ 0.0082\pm  0.0028\pm 0.0022$&$ 0.69\pm 0.23\pm 0.15$\\  \hline
  0.00025  &  0.0004   &$ 0.00032$&$ 0.052$&$0.54$&$ 0.0061\pm  0.0028\pm 0.0027$&$ 0.49\pm 0.22\pm 0.21$\\  \hline
  0.0004   &  0.00063  &$ 0.00050$&$ 0.082$&$0.55$&$ 0.0133\pm  0.0029\pm 0.0036$&$ 0.96\pm 0.21\pm 0.24$\\  \hline
  0.00063  &  0.001    &$ 0.00079$&$  0.13$&$0.55$&$ 0.0172\pm  0.0032\pm 0.0031$&$ 1.08\pm 0.20\pm 0.16$\\  \hline
  0.001    &  0.0016   &$  0.0013$&$  0.21$&$0.55$&$ 0.0125\pm  0.0035\pm 0.0030$&$ 0.40\pm 0.11\pm 0.15$\\  \hline
  0.0016   &  0.0025   &$  0.0020$&$  0.33$&$0.55$&$ 0.0125\pm  0.0040\pm 0.0027$&$ 0.36\pm 0.12\pm 0.10$\\  \hline
  0.0025   &  0.004    &$  0.0031$&$  0.52$&$0.55$&$ 0.0150\pm  0.0048\pm 0.0049$&$ 0.36\pm 0.11\pm 0.12$\\  \hline
  0.004    &  0.0063   &$  0.0049$&$  0.66$&$0.46$&$ 0.0187\pm  0.0067\pm 0.0067$&$ 0.32\pm 0.12\pm 0.12$\\  \hline
  0.0063   &  0.01     &$  0.0077$&$  0.69$&$0.30$&$  0.048\pm  0.011\pm  0.008$&$ 0.52\pm 0.12\pm 0.10$\\  \hline
  0.01     &  0.016    &$   0.012$&$  0.74$&$0.21$&$  0.040\pm  0.019\pm  0.016$&$ 0.29\pm 0.14\pm 0.12$\\  \hline
  0.016    &  0.025    &$   0.019$&$  0.81$&$0.14$&$  0.037\pm  0.035\pm  0.024$&$ 0.19\pm 0.18\pm 0.12$\\  \hline
  0.025    &  0.04     &$   0.028$&$  0.91$&$0.11$&$   0.03\pm  0.11\pm   0.13$&$ 0.12\pm 0.40\pm 0.48$\\
\hline \hline
\multicolumn{7}{|c|}{200~GeV data}\\
\hline
  0.00004  &  0.000063 &$0.000051$&$0.0091$&$0.46$&$ 0.0073\pm  0.0043\pm 0.0044$&$ 0.74\pm 0.44\pm 0.40$\\  \hline
  0.000063 &  0.0001   &$0.000081$&$ 0.016$&$0.51$&$ 0.0053\pm  0.0036\pm 0.0023$&$ 0.58\pm 0.39\pm 0.22$\\  \hline
  0.0001   &  0.00016  &$ 0.00013$&$ 0.026$&$0.54$&$ 0.0057\pm  0.0033\pm 0.0021$&$ 0.64\pm 0.37\pm 0.22$\\  \hline
  0.00016  &  0.00025  &$ 0.00020$&$ 0.043$&$0.57$&$ 0.0042\pm  0.0032\pm 0.0019$&$ 0.46\pm 0.35\pm 0.21$\\  \hline
  0.00025  &  0.0004   &$ 0.00032$&$ 0.070$&$0.58$&$ 0.0039\pm  0.0033\pm 0.0025$&$ 0.39\pm 0.33\pm 0.25$\\  \hline
  0.0004   &  0.00063  &$ 0.00050$&$  0.11$&$0.58$&$ 0.0095\pm  0.0034\pm 0.0024$&$ 0.86\pm 0.31\pm 0.20$\\  \hline
  0.00063  &  0.001    &$ 0.00079$&$  0.17$&$0.58$&$ 0.0111\pm  0.0038\pm 0.0028$&$ 0.83\pm 0.29\pm 0.20$\\  \hline
  0.001    &  0.0016   &$  0.0013$&$  0.28$&$0.59$&$ 0.0104\pm  0.0042\pm 0.0027$&$ 0.45\pm 0.18\pm 0.14$\\  \hline
  0.0016   &  0.0025   &$  0.0020$&$  0.44$&$0.59$&$-0.0002\pm  0.0049\pm 0.0037$&$-0.01\pm 0.17\pm 0.13$\\  \hline
  0.0025   &  0.004    &$  0.0031$&$  0.65$&$0.56$&$ 0.0196\pm  0.0061\pm 0.0038$&$ 0.54\pm 0.17\pm 0.11$\\  \hline
  0.004    &  0.0063   &$  0.0048$&$  0.71$&$0.39$&$  0.048\pm  0.011\pm  0.009$&$ 0.85\pm 0.19\pm 0.16$\\  \hline
  0.0063   &  0.01     &$  0.0077$&$  0.76$&$0.26$&$  0.014\pm  0.018\pm  0.010$&$ 0.16\pm 0.21\pm 0.12$\\  \hline
  0.01     &  0.016    &$   0.012$&$  0.80$&$0.18$&$  0.099\pm  0.034\pm  0.025$&$ 0.75\pm 0.26\pm 0.19$\\  \hline
  0.016    &  0.025    &$   0.018$&$  0.87$&$0.13$&$  0.010\pm  0.076\pm  0.046$&$ 0.05\pm 0.40\pm 0.24$\\  \hline
  0.025    &  0.04     &$   0.026$&$  0.98$&$0.10$&$  -0.43\pm  0.81\pm   0.61$&$ -1.7\pm  3.2\pm  2.4$\\
\hline \hline
\end{tabular}
}
\end{center}
\end{table}
\begin{table}[h!]
\begin{center}
{\footnotesize
\caption{\small
Values of $A_1^{\rm p}$ and $g_1^{\rm p}$ with their statistical and systematic uncertainities
 as a function of $x$ and the average values of $x$, $Q^2$ and $y$, shown for
the combination of  160~GeV and 200~GeV data.
The maximum value of $Q^2$ used in the analysis is $1\,(\GeV/c)^2$. 
Bins in $x$ are of equal width
in log$_{10}x$.}
\label{tab:a1x_g1_combined}
\begin{tabular}{|r@{--}l|c|c|c|c|c|}
\hline \hline
\multicolumn {2}{|c|}{}&&&&& \\
\multicolumn {2}{|c|}{$x$ range} & $\langle x \rangle$ & $\langle Q^2 \rangle $  & $\langle y\rangle $  & \multicolumn{1}{c|}{$A_1^{\rm p}$} & \multicolumn{1}{c|}{$g_1^{\rm p}$} \\
\multicolumn {2}{|c|}{}   &  &   [(GeV$/c$)$^2$]&  &  &  \\
\hline \hline
  0.00004  &  0.000063 &$0.000052$&$0.0076$&$0.43$&$ 0.0073\pm  0.0030\pm 0.0034$&$  0.62\pm  0.25\pm  0.22$\\  \hline
  0.000063 &  0.0001   &$0.000081$&$ 0.013$&$0.48$&$ 0.0064\pm  0.0025\pm 0.0025$&$  0.58\pm  0.22\pm  0.19$\\  \hline
  0.0001   &  0.00016  &$ 0.00013$&$ 0.022$&$0.52$&$ 0.0079\pm  0.0022\pm 0.0019$&$  0.73\pm  0.20\pm  0.12$\\  \hline
  0.00016  &  0.00025  &$ 0.00020$&$ 0.037$&$0.54$&$ 0.0065\pm  0.0021\pm 0.0019$&$  0.59\pm  0.19\pm  0.15$\\  \hline
  0.00025  &  0.0004   &$ 0.00032$&$ 0.059$&$0.56$&$ 0.0052\pm  0.0021\pm 0.0021$&$  0.45\pm  0.18\pm  0.18$\\  \hline
  0.0004   &  0.00063  &$ 0.00050$&$ 0.094$&$0.56$&$ 0.0117\pm  0.0022\pm 0.0021$&$  0.92\pm  0.17\pm  0.14$\\  \hline
  0.00063  &  0.001    &$ 0.00079$&$  0.15$&$0.56$&$ 0.0147\pm  0.0024\pm 0.0026$&$  0.98\pm  0.16\pm  0.15$\\  \hline
  0.001    &  0.0016   &$  0.0013$&$  0.24$&$0.56$&$ 0.0117\pm  0.0027\pm 0.0028$&$  0.42\pm  0.09\pm  0.14$\\  \hline
  0.0016   &  0.0025   &$  0.0020$&$  0.38$&$0.57$&$ 0.0073\pm  0.0031\pm 0.0016$&$ 0.212\pm 0.098\pm 0.053$\\  \hline
  0.0025   &  0.004    &$  0.0031$&$  0.57$&$0.55$&$ 0.0167\pm  0.0037\pm 0.0024$&$ 0.426\pm 0.094\pm 0.070$\\  \hline
  0.004    &  0.0063   &$  0.0049$&$  0.67$&$0.44$&$ 0.0273\pm  0.0057\pm 0.0040$&$ 0.476\pm 0.098\pm 0.077$\\  \hline
  0.0063   &  0.01     &$  0.0077$&$  0.71$&$0.29$&$ 0.0386\pm  0.0095\pm 0.0054$&$  0.42\pm  0.11\pm  0.07$\\  \hline
  0.01     &  0.016    &$   0.012$&$  0.76$&$0.20$&$  0.054\pm  0.016\pm  0.012$&$  0.40\pm  0.12\pm  0.09$\\  \hline
  0.016    &  0.025    &$   0.019$&$  0.82$&$0.14$&$  0.033\pm  0.032\pm  0.017$&$  0.16\pm  0.16\pm  0.09$\\  \hline
  0.025    &  0.04     &$   0.028$&$  0.91$&$0.11$&$   0.02\pm  0.11\pm   0.09$&$  0.09\pm  0.40\pm  0.32$\\
\hline \hline
\end{tabular}
}
\end{center}
\end{table}

\end{document}